\documentclass[reprint,amsmath,amssymb,pra,longbibliography]{revtex4-2}
\usepackage{graphicx}
\usepackage{color}
\usepackage{dcolumn}
\usepackage{bm}

\usepackage[caption=false]{subfig}

\newcommand{\Par}{\partial}
\newcommand{\vare}{\varepsilon}
\newcommand{\abs}[1]{\left| #1 \right|}

\newcommand{\cbrc}[1]{\left( #1 \right)}

\newcommand{\INF}{\infty}
\newcommand{\RE}{\mathrm{Re}}
\newcommand{\IM}{\mathrm{Im}}
\newcommand{\conj}[1]{\overline{#1}}

\begin{document}


\title{Frequency perturbation theory of bound states in the continuum\\
  in a periodic waveguide}

\author{Amgad Abdrabou}
\email{Corresponding author: abdrabou@zju.edu.cn}
\affiliation{School of Mathematical Sciences, Zhejiang University,
  Hangzhou, Zhejiang 310027, China}

\author{Ya Yan Lu}
\affiliation{Department of Mathematics, City University of Hong Kong,
  Kowloon, Hong Kong, China} 
 
\date{\today}

\begin{abstract}
In a lossless periodic structure,  a bound
state in the continuum (BIC) is characterized  by a real  frequency
and a real Bloch wavevector  for which there exist waves propagating to or from
infinity  in the surrounding media. For applications, it is important
to analyze the high-$Q$ resonances that either exist naturally for wavevectors near
that of the BIC or appear when the structure is perturbed. Existing theories provide
quantitative results  for 
the complex frequency  (and the $Q$-factor) of resonant modes that
appear/exist  due to structural perturbations or wavevector
variations. When a periodic structure is regarded as a 
periodic waveguide, eigenmodes are often analyzed for a given real
frequency. In this paper, we consider 
periodic waveguides with a BIC, and study the eigenmodes for a given
real frequency near the frequency of the BIC. It turns out that 
such eigenmodes near the BIC always
have a complex Bloch wavenumber, but they may or may not be 
leaky modes that radiate out power laterally to infinity. These eigenmodes can
also be \textcolor{black}{the so-called} {\it complex modes} that decay exponentially in the lateral
direction. Our study is relevant for applications of BICs in
\textcolor{black}{periodic} optical 
waveguides, and it is also helpful for analyzing photonic devices
operating near the frequency of a BIC. 
\end{abstract}

\maketitle

\section{Introduction}
\label{S1}


In recent years, bound states in the continuum (BICs) have been the
central topic of many studies in
photonics~\cite{hsu16,kosh19,azzam,sad21}. 
For a structure with at least one open spatial direction, a photonic BIC is an eigenmode of
the governing Maxwell's equations satisfying two conditions: (1) it
decays rapidly in the open spatial direction, and (2) at the
same frequency as the BIC, there exist waves that propagate to or from infinity
in the open spatial direction. For a periodic structure 
sandwiched between two homogeneous media, such as a photonic crystal
slab~\cite{padd00,ochiai01,tikh02,shipman07,lee12,hsu13,yang14,gan16,fudan,notomi} or a periodic array of cylinders~\cite{shipman03,port05,mari08,bulg14,hu15,yuan17,hu20pra,amgad21}, a BIC is characterized by
its frequency and Bloch wavevector, the direction perpendicular to the
periodic layer is the open spatial direction, and propagating
diffraction orders compatible with the BIC frequency and wavevector
are the waves that propagate to or from infinity.
\textcolor{black}{For optical waveguides with an invariant
  direction~\cite{zou15,gomis17,muk18,lijun21oe}, a 
  BIC is characterized by its frequency and propagation constant.}


Most applications of BICs are related to high-$Q$ resonances that 
exist near a BIC or appear when a BIC is destroyed. In a periodic structure, a
resonant mode is an outgoing solution of the Maxwell's equations with a real
Bloch wavevector and a complex frequency~\cite{fan02,link19}. A high-$Q$ resonance leads to
local field enhancement~\cite{yoon15,moca15,bulg17,hu20_1,lfe22} and
sharp features in scattering spectra \cite{shipman05,gipp05,shipman12,bykov15,blan16,wu22} that
are useful for lasing, sensing, switching, nonlinear optics, etc. To
obtain a high-$Q$ resonance, the standard way is to perturb the
structure~\cite{kosh18,hu18,perturb20}. Actually, a structural perturbation does
not always destroy a BIC. If the BIC is protected by a symmetry, it
continues to exist when the structure is perturbed preserving the
symmetry. Some BICs are not  protected by symmetry in the 
sense of symmetry mismatch, but can nevertheless persist
under certain structural
perturbations~\cite{zhen14,bulg17pra,yuan17_4,robust21,robustoe}. In
general, if a structural perturbation contains a sufficient number of 
parameters, a generic BIC can survive the perturbation if the
parameters are properly tuned~\cite{para20,para21}. On the other hand,
high-$Q$ resonant modes naturally exist near a BIC in a periodic
structure without any structural perturbation. In fact, a BIC is a
special point in a band of resonant modes that depend on the Bloch wavevector continuously. For a
lossless structure, the $Q$ factor of the resonant mode tends to
infinity as its wavevector tends to that of the BIC.
The asymptotic relation between the $Q$ factor and wavevector
difference  can be determined using a perturbation
method~\cite{bistab17,perturb18,perturb20}. It is known that for some
special BICs, the $Q$ factor of the 
nearby resonant mode tends to infinity extremely quickly~\cite{perturb20,perturb18,jin19}.

A periodic structure sandwiched between two homogeneous media can  
be considered as a periodic waveguide. Eigenmodes in optical
waveguides are often analyzed for a given real frequency. 
In this paper, we study eigenmodes of a periodic waveguide for 
frequencies near the frequency of a BIC. For simplicity, we consider 
two-dimensional (2D) structures with a single periodic direction,  and study only
eigenmodes in the $E$ polarization. 
At a real frequency, a waveguide mode is either a guided mode that
decays exponentially in the lateral direction or 
a leaky mode that radiates out
power to infinity (also in the lateral direction). In the case of a periodic waveguide (with a
periodicity along the waveguide axis), the propagation constant is the
Bloch wavenumber in the periodic direction. For a lossless waveguide,
regular guided modes below the light line
have a real propagation constant and form bands that depend on the
frequency continuously. A BIC is also a guided mode, but it lies above
the light line and is \textcolor{black}{usually} an isolate point in the real
wavenumber-frequency plane. For open lossless periodic waveguides,
there exist guided
modes with a complex propagation constant and they are the so-called
{\it complex modes}~\cite{cmode20}.
\textcolor{black}{A {\it complex mode}, like a complex eigenvalue of a
  real nonsymmetric matrix, exists because
  the  periodic-waveguide eigenvalue problem for a given frequency is
  not self-adjoint.}
{\it Complex modes} are 
well-known for waveguides with shielded boundaries~\cite{mrozo97}, but
they also exist in open lossless dielectric
waveguides~\cite{jablo94,xie11,cmode21}.
\textcolor{black}{It should be emphasized that the complex propagation
  constant of a {\it complex mode} is not caused by material or
  radiation loss, and a {\it complex mode}   is still a
  guided mode, since it decays  exponentially in the lateral
  direction. A different kind of waveguide
  modes with a complex propagation constants are the well-known 
leaky modes~\cite{snyder,vassallo,aop}.} 
Due to the radiation loss (power is radiated out in the lateral
direction), the propagation constant of a leaky mode is always
complex.
\textcolor{black}{Unlike a {\it complex mode}}, the amplitude of a leaky mode grows exponentially in the
lateral direction. Both {\it complex}  and leaky modes form bands, and
each band is given by the propagation constant being a complex-valued
function of the real frequency.
The purpose of this work is to reveal the connection between BICs and
leaky or {\it complex} modes. Using a perturbation method, we show
that \textcolor{black}{when the frequency is perturbed, a BIC does not
  always become a leaky mode. In fact, it can also become a  {\it complex mode}.}

The rest of this paper is organized as follows. In Sec.~\ref{S2}, we
present a summary and an example for various eigenmodes in a periodic
structure.  In Sec.~\ref{S3}, we use a perturbation method to analyze
the waveguide modes near a BIC.  Numerical
examples are presented in Sec.~\ref{S4} to validate the perturbation
theory. The paper is concluded with some comments in Sec.~\ref{S5}.

\section{Eigenmodes in 2D periodic structures}
\label{S2}

\begin{figure}[htbp!]
        \centering
        \subfloat[]{\includegraphics[width=0.8\linewidth]{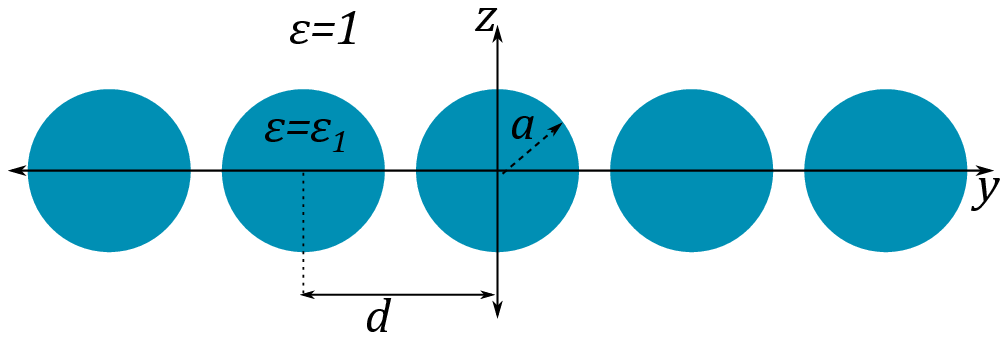}}\\
        \subfloat[]{\includegraphics[width=0.8\linewidth]{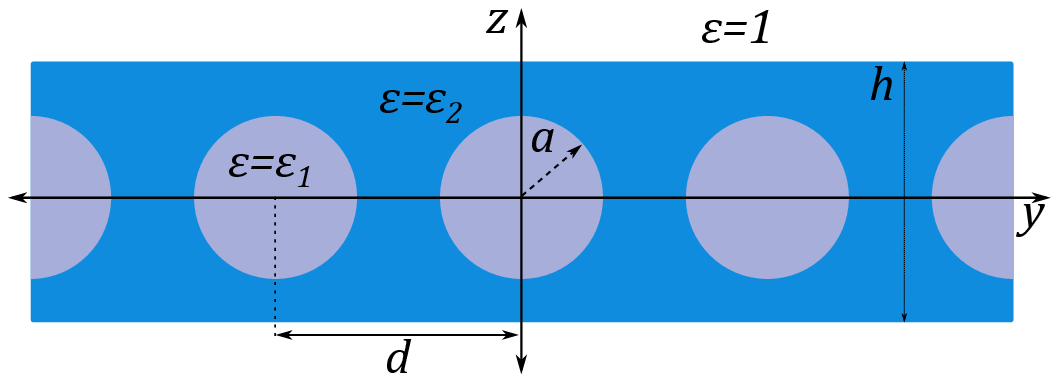}}
        \caption{Schematic diagrams of two periodic structures with period $d$ along the $y$-axis. }
        \label{Fig1}
\end{figure}

In this section, we recall the definitions of various eigenmodes in 2D
periodic structures and illustrate their connections by a numerical
example. Consider a periodic structure that is invariant in $x$,
periodic in $y$ with period $d$, bounded in $z$ by $|z| < h/2$ for
some $h>0$, and 
surrounded by air. The dielectric function $\vare$ 
is a real function of ${\bm r}=(y,z)$ and satisfies
$\vare(y+d,z) = \vare({\bm r})$ for all ${\bm r}$, $\vare({\bm r}) =
1$ for $|z| > h/2$, and $\max \vare({\bm r}) > 1$. 
Two examples are shown in Fig.~\ref{Fig1}.
Panel (a) shows a periodic array of circular cylinders with radius $a$ and
dielectric constant $\vare_1$, and panel (b) depicts a slab of thickness
$h$ and dielectric constant $\vare_2$, containing a periodic array of
cylinders with radius $a$ and dielectric constant $\vare_1$.

For the $E$-polarization, the $x$ component of the time-harmonic
electric field,  denoted as $u$, satisfies the following 2D Helmholtz 
equation:  
\begin{equation}
  \label{Eq1}
\partial_y^2 u +\partial_z^2 u +k^2 \vare({\bm r})u = 0,
\end{equation}
where $k=\omega/c$ is the freespace wavenumber, $\omega$ is the angular
frequency, and $c$ is the speed of light in vacuum, and the time
dependence is $e^{-i   \omega t}$. An eigenmode of such a periodic
structure is a solution of Eq.~\eqref{Eq1} given by 
\begin{equation}
  \label{Eq2}
u({\bm r}) = \phi({\bm r})\,e^{i\beta y},
\end{equation}
where $\beta$ is the Bloch wavenumber satisfying $\abs{\RE(\beta)}\leq
\pi/d$, and $\phi({\bm r})$ is periodic  in $y$ with period
$d$. In the free space given by $\abs{z}> h/2$, 
the eigenmode can  be expanded in plane waves as 
\begin{equation}
  \label{Eq3}
u({\bm r}) = \sum_{m=-\INF}^{\INF} \hat{u}_{m}^{\pm} e^{i(\beta_m y\pm\alpha_m z)}, \quad \pm z > h/2,
\end{equation}
where $\hat{u}_m^\pm$ are the expansion coefficients, $\beta_0 =\beta$,
\begin{equation}
  \label{Eq4}
\beta_m = \beta+\frac{2\pi m}{d},\quad \alpha_m  =\sqrt{k^2-\beta_m^2},
\end{equation}
and the square root is defined using a branch cut along the negative
imaginary axis. 

An eigenmode must satisfy a proper boundary condition as $z \to \pm
\infty$. If $\phi({\bm r}) \to 0$  as $\abs{z} \to \INF$, then the
eigenmode is a guided mode.  If both $\beta$ and $k$ are real, and $ k
< \abs{\beta}$, the guided mode is a regular one below the light
line. The regular guided modes form bands that depend on $\beta$ and $k$
continuously. A BIC is also a guided mode, but it is above the light
line. More precisely, both $\beta$ and $k$ of a BIC are real and $k >
|\beta|$.
Since a BIC must decay as $z \to \pm \infty$, if for any $m$,
$\alpha_m$ is real (note that at least $\alpha_0 > 0$), then
$\hat{u}_m^\pm$ in Eq.~(\ref{Eq3}) must vanish, because they are the
coefficients of propagating plane waves. 
The periodic structure can also support {\it complex modes} which are
guided modes with a complex $\beta$~\cite{cmode20}. Since the
structure is non-absorbing and the field decays to zero as $z\to \pm
\infty$, the {\it complex modes} are unrelated to absorption and radiation
losses. They exist because the eigenvalue problem for a given
frequency (where $\beta$ is the eigenvalue) is not self-adjoint~\cite{jablo94}. The
existence of {\it complex modes} is similar to the existence of complex eigenvalues
for a real non-symmetric matrix.

Eigenmodes can also be defined using an outgoing radiation
condition. In that case, the eigenmode radiates out power to
infinity in the lateral direction, i.e., as $z\to \pm \infty$. A leaky
mode is an eigenmode with a real $k$ and an outgoing wave
field~\cite{snyder,vassallo,aop}. 
Since a leaky mode is losing power as it propagates
forward, $\beta$ should have a positive imaginary part, so that the
amplitude of the mode decays as it propagates forward. On the other
hand, a complex $\beta$ implies that $\mbox{Im}(\alpha_0) < 0$, thus,
the plane waves $\exp [  i (\beta y \pm \alpha_0 
z)]$ blow  up and the field of a leaky mode grows exponentially as $z
\to \pm \infty$. A resonant mode is also an eigenmode satisfying the
outgoing 
radiation condition, but it is given for a real $\beta$~\cite{fan02,link19}. Since
$\beta$ is real, the amplitude is uniform in the $y$ direction, to 
radiate out power to infinity in the lateral direction, a resonant
mode must have a complex frequency (with a negative imaginary part),
so that it decays with time.  This implies that $\mbox{Im}(\alpha_0)$
is also negative, and the field is unbounded as $z \to \pm \infty$. 

\begin{figure}[htbp!]
  \centering 
  \includegraphics[width=\linewidth]{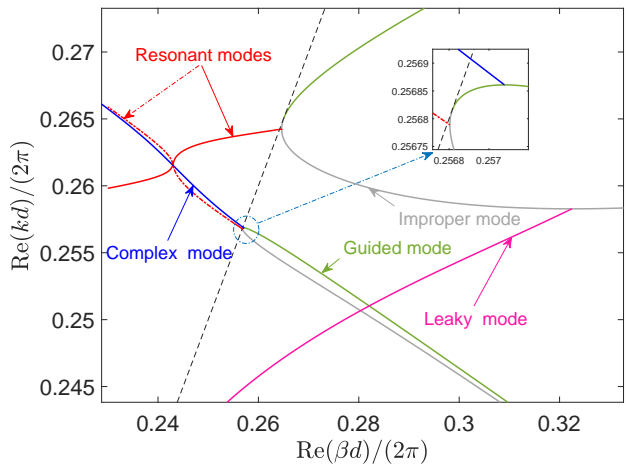}  
  \caption{Normalized real parts of $k$ and $\beta$ for different
    modes near the light line. We show resonant (red), guided (green),
    improper (gray), leaky (purple) and {\it complex} (blue) mode curves.}
  \label{Fig2}
\end{figure}

To illustrate the different eigenmodes, we present an example for the
periodic structure shown  Fig.~\ref{Fig1}(b).
For  $\vare_1 = 1$, $\vare_2 =11.56$, $h = 1.8d$ and $a = 0.25d$, we
calculate the dispersion curves for various eigenmodes
\textcolor{black}{using a numerical method based on a nonlinear
  eigenvalue formulation~\cite{hu15,yuan17,ep18}}. The results are
shown in Fig.~\ref{Fig2}. 
The dispersion curves
for regular guided, leaky, {\it complex},  resonant, and \textcolor{black}{the
  so-called} improper modes are
shown as green, purple, blue, red, and gray curves, respectively.
For resonant and {\it complex}/leaky modes, only the real parts of 
$k$ or $\beta$ are shown in the figure. 
The dashed line is the light line $k = \beta$. 
Two guided modes emerge from the light line tangentially. The
dispersion curve of the lower guided mode has a local maximum where a
{\it complex mode} appears~\cite{cmode20}. An improper mode is a solution 
with a real $k$ and a real $\beta$, but it grows exponentially as $z
\to \pm \infty$. Two improper 
modes emerge at the same points on the light line as the regular guided
modes. A leaky mode appears at the minimum point on the dispersion
curve of an improper mode~\cite{link19}. The resonant modes are connected to the
improper modes where the dispersion curves (of the improper modes) have
an infinite slope~\cite{link19}. At a particular value of $\beta$, the
two resonant modes coalesce and form an exceptional point~\cite{ep18,ep20}. 

\section{Perturbation analysis}
\label{S3}

In this section, we develop a perturbation theory for
\textcolor{black}{waveguide} modes (leaky  or {\it complex} modes) near a BIC in a periodic structure. 
As in Sec.~II, we consider a 2D lossless periodic structure that is
translationally invariant in $x$, periodic in $y$ with period $d$, and
surrounded by air for $|z| > h/2$, and focus on $E$-polarized Bloch
eigenmodes with a real frequency. Suppose the periodic structure
supports a BIC
$u_*({\bm r}) = \phi_*({\bm r}) e^{ i \beta_* y}$
with Bloch wavenumber $\beta_*$ and frequency
$\omega_*$ (freespace wavenumber $k_*=\omega_*/c$), we assume $k_*$
satisfies
\begin{equation}
  \label{chan1}
  |\beta_*| < k_* < \frac{2\pi}{d} - |\beta_*|,   
\end{equation}
then $\alpha_* = \sqrt{ k_*^2 - \beta_*^2}$ is positive,  and for $m\ne 0$, 
$\alpha_m^* = [  k_*^2 - (\beta_* +
2\pi m/d)^2]^{1/2}$ is pure imaginary with a positive imaginary
  part. This means that for the pair $\{ \beta_*, k_* \}$, there is
  only one radiation channel for positive or negative $z$,
  respectively. Now, for a given real $k$ near $k_*$, we seek 
  a Bloch eigenmode
  $u({\bm r}) = \phi({\bm r}) e^{i \beta y}$ that either decays exponentially
  or radiates out power as $z \to \pm \infty$.
In terms of $\phi$, Eq.~\eqref{Eq1} takes the form 
\begin{equation}\label{Eqq1}
\partial_y^2 \phi  + \partial_z^2 \phi + 2i\beta \Par_y\phi +
[ k^2\vare({\bm r})-\beta^2 ] \phi = 0.  
\end{equation}
Since \textcolor{black}{the periodic structure is embedded in a
  homogeneous medium}, a BIC is an isolated point in the real $\beta$-$k$
  plane (when $d$ is the true minimum period of the structure), if $k \ne k_*$, $\beta$ is always complex. To find the Bloch
  mode with a complex $\beta$, we use a perturbation method assuming 
  $|(\omega-\omega_*)/\omega_*| = |(k-k_*)/k_*|$ is small. For simplicity, we let $\delta = k^2
  - k_*^2$ and expand $\beta$ and $\phi$ in power series of
  $\delta$. It turns out that we need to use power series of
  $\sqrt{|\delta|}$ when the BIC carries zero power.

\subsection{BIC with nonzero power}  

For $\delta \ne 0$, we seek 
$\beta$ and $\phi$ from the following power series:
\begin{eqnarray}
\beta &=& \beta_* + \beta_1 \delta + \beta_2 \delta ^2 + \cdots, \\
\phi  &=& \phi_* + \phi_1 \delta + \phi_2 \delta ^2 + \cdots. 
\end{eqnarray}
Inserting the above into Eq.~\eqref{Eqq1} and comparing terms of equal
powers of $\delta$,  we obtain  
\begin{eqnarray}
\label{Eqq2}   
  \mathcal{O}(1):  && \quad \mathcal{L}\phi_* = 0,  \\
\label{Eqq3}    
\mathcal{O}(\delta): && \quad  \mathcal{L} \phi_1 = 2\beta_1 (\beta_* \phi_*
                        -i \Par_y\phi_*) -\vare({\bm r}) \phi_*,  \\
  \label{Eqq4}
\mathcal{O}(\delta^2):  && \quad \mathcal{L} \phi_2 = 2\beta_1
                           (\beta_* \phi_1 - i \Par_y\phi_1)
                           -\vare({\bm r}) \phi_1 \\
&&  \nonumber \hspace{1cm} +2\beta_2 (  \beta_* \phi_* -i \Par_y\phi_*
   ) + \beta_1^2 \phi_*,
\end{eqnarray}
where $\mathcal{L} \equiv \partial_y^2 + \partial_z^2
+2i\beta_*\Par_y+  k_*^2\vare-\beta_*^2$.

Equation~(\ref{Eqq2}) is simply the governing equation of the
BIC. The inhomogeneous equations (\ref{Eqq3}) and (\ref{Eqq4}) are
singular and have no solution unless the right hand sides are
orthogonal to $\phi_*$. Let $\Omega$ be the domain given by $0 <y <
d$ and $-\infty <z <\INF$. Multiplying $\conj{\phi}_*$ to both sides of
Eq.~(\ref{Eqq3}) and integrating on $\Omega$,
we obtain 
\begin{equation}
  \label{Eqq5}
\beta_1 = \frac{1}{\mathcal{P}} \int_{\Omega} \vare \abs{\phi_*}^2 
d{\bm r}, 
\end{equation}
where
\begin{equation}
  \label{defP}
  \mathcal{P} = -2i \int_\Omega  \conj{u}_*  \frac{\partial 
      u_*}{\partial y}  d{\bm r},  
  \end{equation}
  and it is assumed to be nonzero. In Appendix, we show that $\mathcal{P}$
  is real and proportional to the power carried by the BIC  in the $y$
  direction. Since we assume the BIC carries a nonzero power,
  $\mathcal{P} \ne 0$.  It is clear that $\beta_1$ is real. In
  addition, we note that
  \[
    \beta_1 = \left. \frac{d \beta}{d k^2} \right|_{k=k_*} =
    \frac{1}{2k_*} \left. \frac{      d\beta}{dk} \right|_{k=k_*}.
  \]
  Thus, the slope of the dispersion curve at the BIC point is related to
  $\beta_1$. 
  
  To reveal the nature of this \textcolor{black}{eigenmode}, it is necessary 
to find the first term with a nonzero imaginary part in the power
series of $\beta$. It is possible to write down a formula for $\beta_2$, but it is given in
terms $\phi_1$ which satisfies Eq.~(\ref{Eqq3}). In Appendix, we
show that the imaginary part of 
$\beta_2$ can be expressed (without involving $\phi_1$) as
\begin{equation}
  \label{beta2new}
  \IM(\beta_2) = \frac{ \abs{F_1}^2+\abs{F_2}^2 }{4 d \alpha_*  \mathcal{P}  }
\end{equation}
where $F_1$ and $F_2$ are given by
\begin{equation}
  \label{defF}
  F_j = \int_\Omega \conj{\psi}_j({\bm r})  G({\bm r}) \, d{\bm r}, \quad j=1, 2,   
\end{equation}
$G({\bm r})  =  -2 i \beta_1 \partial_y \phi_*  + [ 2 \beta_* \beta_1 
-\varepsilon({\bm r}) ]  \phi_* $ is the right hand side of Eq.~(\ref{Eqq3}),
$\psi_1$ and $\psi_2$ are related to $w_1$ and $w_2$ by
\begin{equation}
  \label{defpsi}
  w_j({\bm r}) =\psi_j({\bm r}) e^{i \beta_* y}, \quad j=1, 2,   
\end{equation}
$w_1$ and $w_2$ are diffraction solutions of Eq.~(\ref{Eq1}) (with $k$ replaced by
$k_*$) corresponding to incident waves $\exp [  i (\beta_* y \pm \alpha_*
  z) ] $ given for $ z  < -  h/2$ and $z > h/2$, respectively.

We assume the BIC is generic in the sense that $(F_1, F_2) \ne
(0,0)$. Since  $\beta_1$ is real and $\IM(\beta_2) \ne 0$, we have
\begin{equation}
  \label{scaling1}
  \IM(\beta) = \mathcal{O}(\delta^2) = \mathcal{O}(
  |\omega-\omega_*|^2). 
\end{equation}
If $\mathcal{P}$ is positive, then $\beta_1$ is positive, 
the imaginary part of $\beta$ is positive, $\alpha_0 = \sqrt{k^2 -
  \beta^2}$ has a negative imaginary part, the plane wave $e^{i (\beta 
  y + \alpha_0 z)}$ grows exponentially as $z \to +\infty$, and the 
eigenmode is a leaky mode. On the other hand, if
$\mathcal{P} <0$, then $\beta_1 < 0$, $\IM(\beta) < 0$,
$\IM( \alpha_0) > 0$, the plane wave $e^{i (\beta 
  y + \alpha_0 z)}$ decays exponentially as $z \to +\infty$, and the 
eigenmode is a {\it complex mode}. Therefore, if a BIC has a nonzero power,
it is a special point on the dispersion curve for a band of \textcolor{black}{eigenmodes
with a complex $\beta$}. If the power of the BIC is positive, then the
dispersion curve has a positive slope at the BIC point and the eigenmodes are
leaky modes. If the power of the BIC is negative, then the dispersion curve
has a negative slope at the BIC point and the \textcolor{black}{eigenmodes}
are {\it complex modes}. If we assume $\beta_* > 0$, the BIC with a negative
power is a backward wave.

\subsection{BIC with zero power}

If the BIC carries no power in the $y$ direction, the
perturbation method based on power 
series of $\delta$ fails. For a typical standing wave with $\beta_*=0$, the
power is indeed zero. Therefore, it is important to analyze this
special case. To find the eigenmodes near a BIC with a zero power, we try power series in
$\sqrt{|\delta|}$. It is convenient to introduce an
integer $s$, such that $s=1$ if $\delta>0$ and $s=-1$ if $\delta < 0$, 
and expand $\beta$ and $\phi$ as  
\begin{align}
\beta = \beta_* + \beta_1 \sqrt{s\delta} + \beta_2 \delta   + \cdots,\\ 
\phi= \phi_* + \phi_1 \sqrt{s\delta} + \phi_2 \delta + \cdots.
\end{align}
Inserting the above expansions into Eq.~(\ref{Eqq1}) and collecting
terms at the same order, we obtain the following equations for $\phi_*$,
$\phi_1$ and $\phi_2$: 
\begin{align}
\mathcal{O}(1):&\ \mathcal{L} \phi_* = 0,&\label{Eqq9}\\
\mathcal{O}(\sqrt{|\delta|}):&\ \mathcal{L}  \phi_1 =
                             2 \beta_1 ( \beta_* \phi_* -i \Par_y\phi_*), &\label{Eqq10}\\
\mathcal{O}(\delta ):&\ \mathcal{L}  \phi_2 =  2s\beta_1 ( \beta_*
                       \phi_1 - i \Par_y\phi_1) + s \beta_1^2 \phi_*  & \nonumber \\
  & +  2 \beta_2 ( \beta_* \phi_*     - i   \Par_y\phi_*)  - \vare({\bm r}) \phi_*. 
                                          &\label{Eqq11}
\end{align}
Since the power of the BIC is zero, the right hand side of
Eq.~(\ref{Eqq10}) is orthogonal to $\phi_*$, thus $\beta_1$ cannot be
determined from the solvability condition of $\phi_1$. To remove
the unknown $\beta_1$, we 
define $\hat{\phi}_1$ such that $\phi_1 = \beta_1 \hat{\phi}_1$, then
$\hat{\phi}_1$ satisfies 
\begin{equation}
  \label{hatphi1}
\mathcal{L}  \hat{\phi} _1 = G({\bm r}) = 2  \beta_* \phi_* -2 i \partial_y \phi_*
\end{equation}

Multiplying $\conj{\phi}_*$ to both sides of Eq.~(\ref{Eqq11}),
replacing $\phi_1$ by $\beta_1 \hat{\phi}_1$, and integrating on
$\Omega$, we get 
\begin{equation}
  \label{Eqq12}
  s \beta_1^2 \int_{\Omega} \left[
        |\phi_*|^2 +   R({\bm r})  \right]  d{\bm r}
  = \int_\Omega \vare({\bm r}) 
  |\phi_*|^2 d{\bm r}, 
\end{equation}
where 
$R({\bf r}) =  2 \conj{\phi}_* (   \beta_*   \hat{\phi}_1 - i
\partial_y \hat{\phi}_1 )$.
Multiplying Eq.~\eqref{Eqq12} by $\conj{\beta}_1^2$, and comparing the 
imaginary parts of both sides,  we obtain
\begin{equation}
  \label{Eqq17}
  \IM\cbrc{\conj{\beta}_1^2}
=  s \abs{\beta_1}^4 \frac{ \IM \int_\Omega R({\bm r}) d{\bm r} }{\int_{\Omega} \vare \abs{\phi_*}^2 
  d{\bm r}}. 
\end{equation}
In Appendix, we show that
\begin{equation}
  \label{rtof}
  \IM \int_\Omega R({\bm r}) d{\bm r}
  = -\frac{  |F_1|^2 + |F_2|^2 }{4 d \alpha_*}, 
\end{equation}
where $F_1$ and $F_2$ are defined as in Eq.~(\ref{defF}) with a new
  $G({\bm r})$ given in Eq.~(\ref{hatphi1}). This leads to 
\begin{equation}
  \label{imb1s}
  \IM \left( \beta_1^2 \right) 
  =  \frac{  s \abs{\beta_1}^4 \left(  |F_1|^2 + |F_2|^2 \right)} 
{4d\alpha_* \int_{\Omega} \vare \abs{\phi_*}^2  d{\bm r}}.
\end{equation}
Therefore, if the BIC satisfies the condition $(F_1, F_2) \ne (0,0)$, 
then $\beta_1$ has  a nonzero imaginary part and
\begin{equation}
  \label{scaling2}
  \IM(\beta) = \mathcal{O}( \sqrt{ |\delta| }) =
  \mathcal{O} ( |\omega - \omega_*|^{1/2}).
\end{equation}

For $k > k_*$, i.e., $s=1$, $\IM \left( \beta_1^2 \right) $ is
positive, thus $\beta_1$ is  in the first or third quadrant of the complex
plane. It is clear that Eq.~(\ref{Eqq12}) has two solutions for
$\beta_1$. Let these two solutions be $\beta_1^{(1)}$ and $\beta_1^{(2)}$, where
$\beta_1^{(1)}$ is in the first quadrant and  $\beta_1^{(2)}= -
\beta_1^{(1)}$ is in the third quadrant.
If $\beta_*$ of the BIC is positive, then
the mode corresponding to 
$\beta_1^{(1)}$ has a positive $\mbox{Re}(\beta)$, a positive 
$\mbox{Im}(\beta)$, and a negative $\mbox{Im}(\alpha_0)$, and it is a leaky  mode;
the mode corresponding to 
$\beta_1^{(2)}$ has a positive $\mbox{Re}(\beta)$, a negative
$\mbox{Im}(\beta)$, a positive $\mbox{Im}(\alpha_0)$, and it is a {\it
  complex mode}. The results are opposite if $\beta_* < 0$.
Since BICs with zero power are usually standing
waves, the most important 
case is $\beta_* = 0$. In that case, the two modes corresponding to 
$\beta_1^{(1)}$ and $\beta_1^{(2)}$ are both leaky modes, and they are
reciprocal to each other. 

If $k  < k_*$, i.e., $s=-1$, then $\IM(\beta_1^2) < 0$, $\beta_1$ is in the
second or fourth quadrant of the complex plane. Let the two solutions
of Eq.~(\ref{Eqq12}) be $\beta_1^{(1)}$ (in the second quadrant)  and
$\beta_1^{(2)} = - \beta_1^{(1)}$ (in the fourth quadrant). For a BIC
with $\beta_* > 0$,  the
mode corresponding to $\beta_1^{(1)}$ has a positive $\RE(\beta)$, a
positive $\IM(\beta)$, a negative $\IM(\alpha_0)$, and it is a leaky
mode; the mode corresponding to $\beta_1^{(2)}$ has a positive
$\RE(\beta)$, a negative $\IM(\beta)$, a positive $\IM(\alpha_0)$, and it
is a {\it complex mode}. The opposite results are obtained for $\beta_* < 0$. For
$\beta_*=0$, the two modes corresponding to $\beta_1^{(1)}$ and
$\beta_1^{(2)}$ are both {\it complex modes}.

\section{Numerical results}
\label{S4}

\begin{figure}[htbp!]
  \centering 
  \subfloat[]{\includegraphics[width=0.5\linewidth]{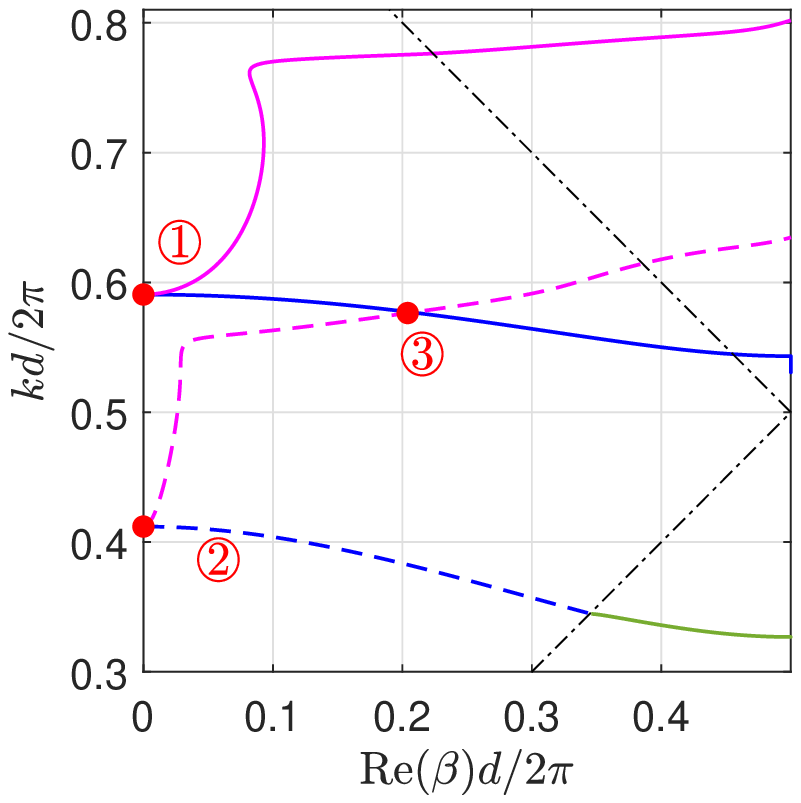}}~
  \subfloat[]{\includegraphics[width=0.5\linewidth]{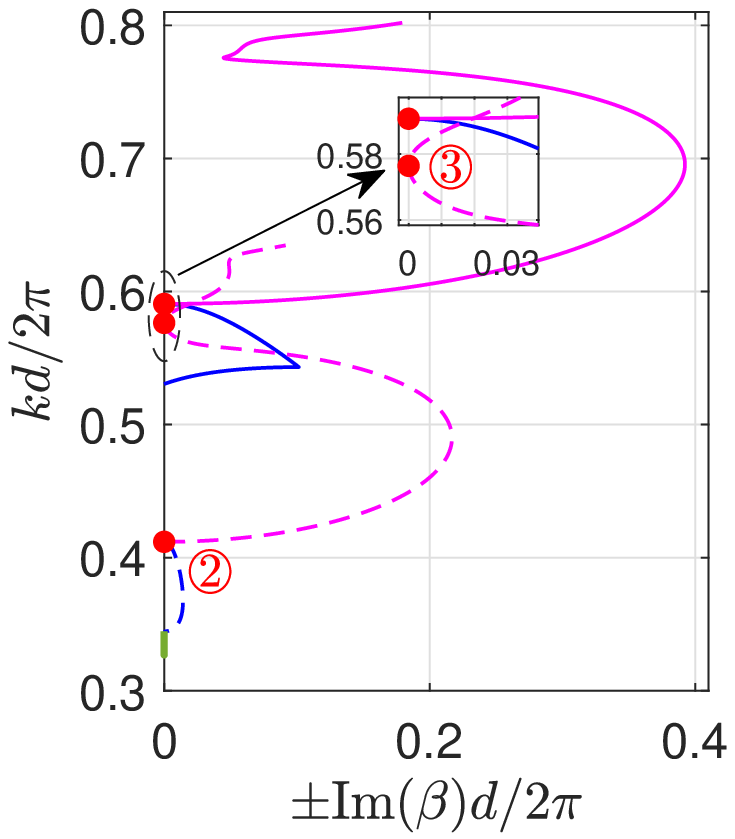}}\\
\subfloat[]{\includegraphics[width=0.5\linewidth]{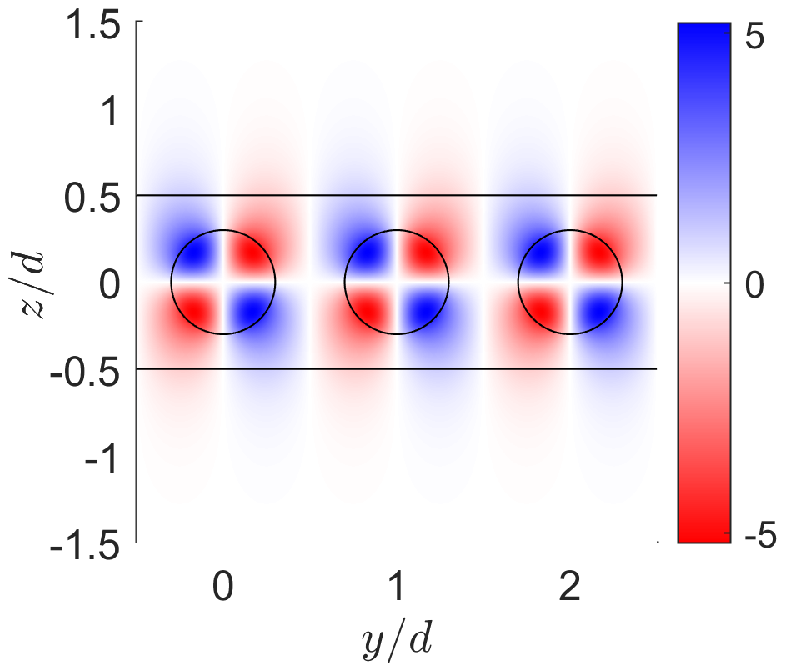}}~
\subfloat[]{\includegraphics[width=0.5\linewidth]{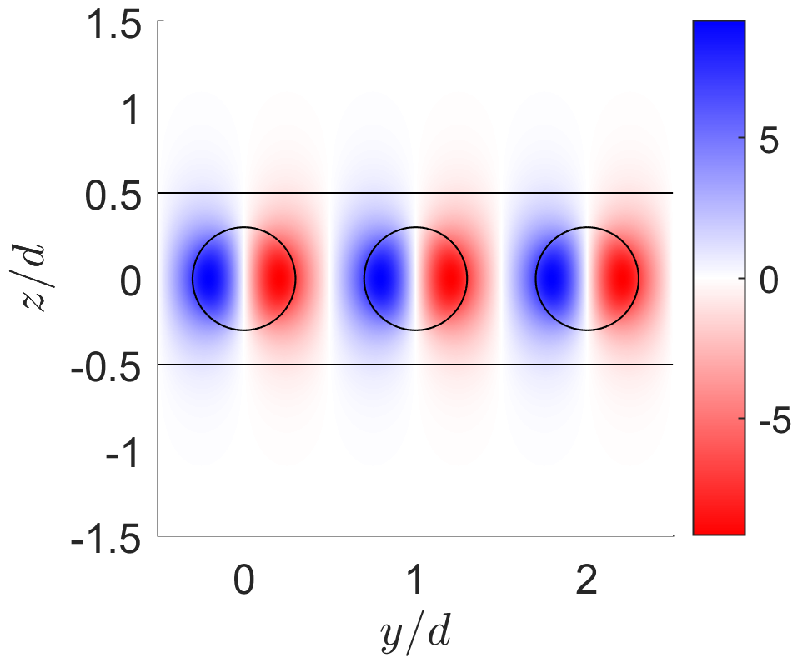}}\\
\subfloat[]{\includegraphics[width=0.5\linewidth]{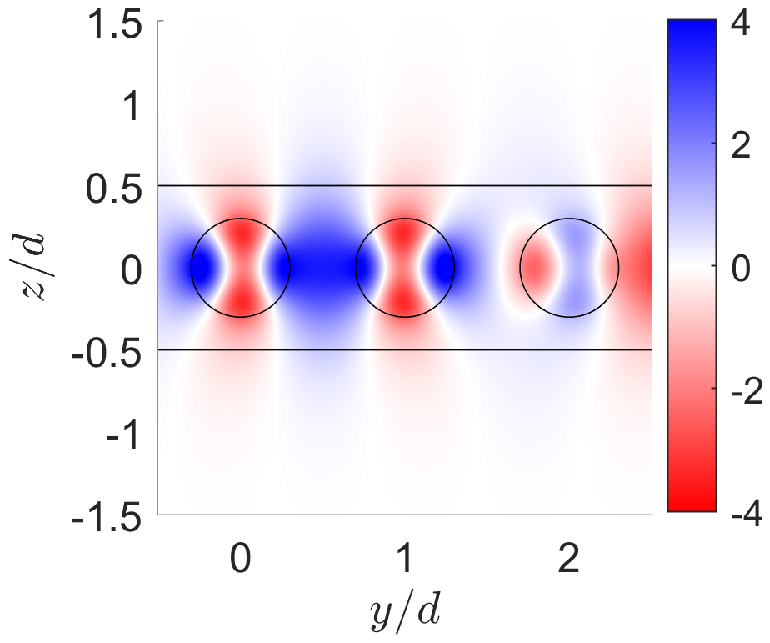}}~
\subfloat[]{\includegraphics[width=0.5\linewidth]{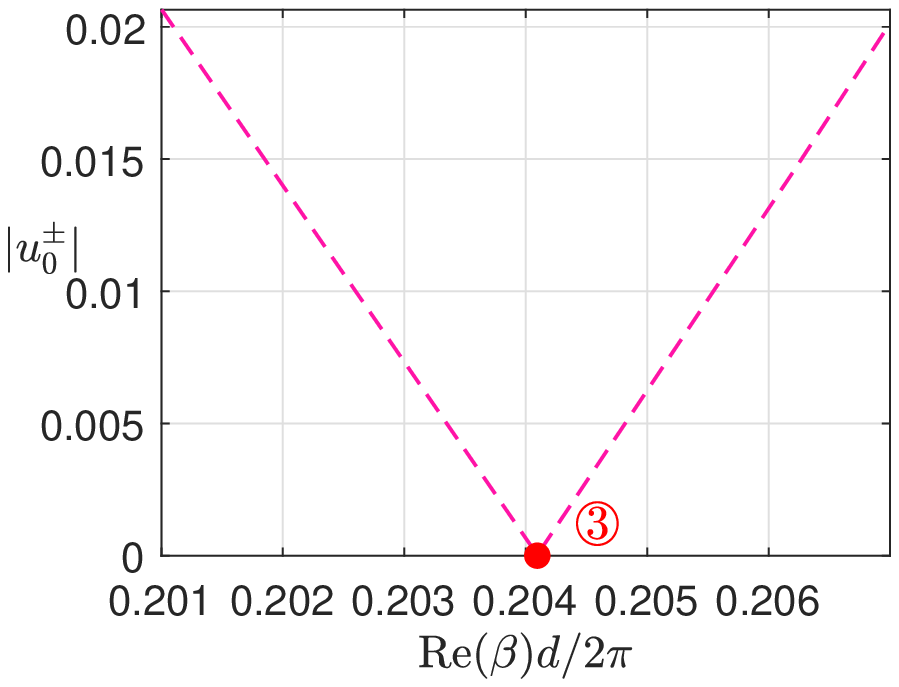}}          
\caption{Example 1: a periodic array of circular cylinders.  (a,b) Dispersion 
  curves of {\it complex} (blue) and leaky (purple) modes, with BICs shown
  as the red dots: (a) 
  $k$  versus $\RE(\beta)$, (b) $k$ versus $\IM(\beta)$. 
  (c-e) Field profiles of BICs: (c) BIC  $\textcircled{\raisebox{-.9pt}1}$, 
  (d) BIC $\textcircled{\raisebox{-.9pt}2}$, 
  (e) BIC $\textcircled{\raisebox{-.9pt}3}$. 
  (f) Coefficient $\hat{u}_0^{\pm}$ of leaky modes near BIC 
  $\textcircled{\raisebox{-0.9pt}3}$. }
  \label{Fig3}
\end{figure}

In this section, we numerically verify the theoretical results
obtained in the previous section.
The first example is a periodic array of circular cylinders
shown in Fig.~\ref{Fig1}(a).
The dielectric constant and the radius 
of the cylinders are  $\vare_1 = 11.56$
and radius $a = 0.3d$, respectively. For the $E$ polarization, the structure 
supports a few BICs. We consider three BICs that are shown as the small red 
dots and marked by  
${\small \textcircled{\raisebox{-.9pt}1}}$, ${\small \textcircled{\raisebox{-.9pt}2}}$ 
and ${\small \textcircled{\raisebox{-.9pt}3}}$ in 
Fig.~\ref{Fig3}(a). 
BICs ${\small \textcircled{\raisebox{-.9pt}1}}$
and ${\small   \textcircled{\raisebox{-.9pt}2}}$
are anti-symmetric standing waves
with $\beta_* = 0$ and their electric fields are odd functions of $y$. The frequencies of
BICs ${\small \textcircled{\raisebox{-.9pt}1}}$ and ${\small   \textcircled{\raisebox{-.9pt}2}}$ are
$\omega_* = 0.5907 (2\pi c/d)$ and
$0.4119 (2\pi c/d)$, respectively, and their field patterns  [real
part of $u_*(y,z)$]  are shown
in Figs.~\ref{Fig3}(c) and \ref{Fig3}(d). BIC ${\small
  \textcircled{\raisebox{-.9pt}3}}$ is a propagating BIC with
$\beta_*
= 0.2041(2\pi /d)$ and $\omega_* = 0.5764 (2\pi c/d)$. The
field pattern of BIC ${\small \textcircled{\raisebox{-.9pt}3}}$ is
quasi-periodic (not periodic) in $y$ and is shown in
Fig.~\ref{Fig3}(e). 

For BICs ${\small \textcircled{\raisebox{-.9pt}1}}$ and
${\small \textcircled{\raisebox{-.9pt}2}}$, 
we found leaky modes for $k>k_*$ and {\it complex modes} for
$k<k_*$, in agreement with the perturbation theory of Sec.~\ref{S3}(B). In
Figs.~\ref{Fig3}(a) and \ref{Fig3}(b), the dispersion curves of the
leaky and {\it complex} modes are shown in purple and blue, respectively.
For each band of leaky or {\it complex} modes, $\beta$ is a complex-valued
function of $k$. The real and imaginary parts of $\beta$ are shown, as
the horizontal axis, in Figs.~\ref{Fig3}(a) and \ref{Fig3}(b),
respectively.
As $k$ is decreased from $k_*$, the
{\it complex mode} emerged from BIC ${\small 
  \textcircled{\raisebox{-.9pt}2}}$
ends below the light line [the
black dashed line with positive slope in Fig.~\ref{Fig3}(a)] at a
local maximum on the dispersion curve of a regular guided
mode~\cite{cmode20}. The solid green curve in Fig.~\ref{Fig3}(a) is the dispersion
curve of the regular guided mode. The
{\it complex mode} emerged from BIC ${\small 
  \textcircled{\raisebox{-.9pt}1}}$ exists up to $\RE(\beta) = \pi/d$,
and turns to a different {\it complex mode} with a fixed $\RE(\beta) =
\pi/d$~\cite{cmode20}. The leaky modes emerged from these two BICs exist
continuously as $k$ is increased and $\RE(\beta)$ passes $\pi/d$ with a finite
derivative $d \beta/dk$. 

On the dispersion curve of the leaky mode emerged from BIC
$ {\small \textcircled{\raisebox{-.9pt}2}}$,
there is a special point
with $\IM(\beta) = 0$, and it is precisely BIC~${\small
  \textcircled{\raisebox{-.9pt}3}}$. Notice that this BIC  is not on
the dispersion curve of the complex mode emerged from BIC~${\small
  \textcircled{\raisebox{-.9pt}1}}$, since $\IM(\beta)$ of the complex
mode at $k_*$ (of BIC~${\small \textcircled{\raisebox{-.9pt}3}}$) is clearly nonzero, as shown in
Fig.~\ref{Fig3}(b). From Fig.~\ref{Fig3}(a), it is clear that
$d\beta/dk > 0$ at $k_*$. This is consistent with the theory developed
in Sec.~III(A). That is, $\beta_1$ is positive and the power of the BIC is positive.
In Fig.~\ref{Fig3}(f), we show the radiation amplitude $\hat{u}_0^\pm$
[defined in Eq.~(\ref{Eq3})] of the leaky mode as a function of
$\beta$. Since $\hat{u}_0^\pm$ depends on the scaling, we assume the
leaky mode satisfies $u(y,h/2)=1$. It is clear that $\hat{u}_0^\pm =0$ for $\beta  = \beta_*$. 
Therefore, as $k \to k_*$, $\IM(\beta)\to 0$, the leaky mode ceases to
decay along the $y$-axis and it stops radiating power in the
transverse direction.

\begin{figure}[htbp!]
  \centering
  \subfloat[]{\includegraphics[width=0.5\linewidth]{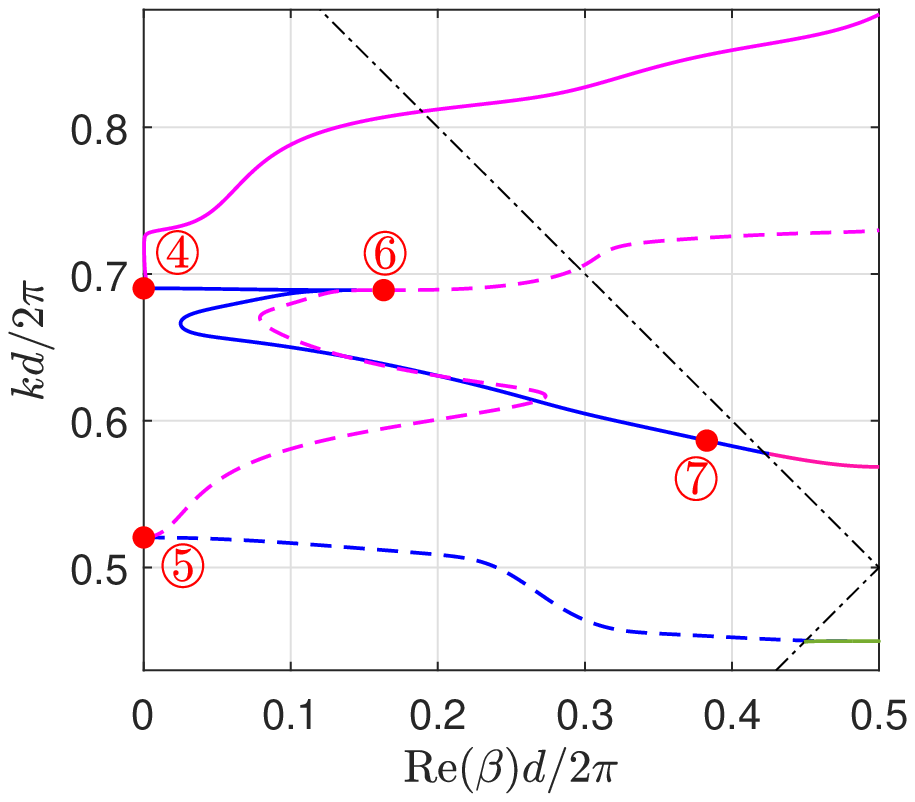}}~
  \subfloat[]{\includegraphics[width=0.5\linewidth]{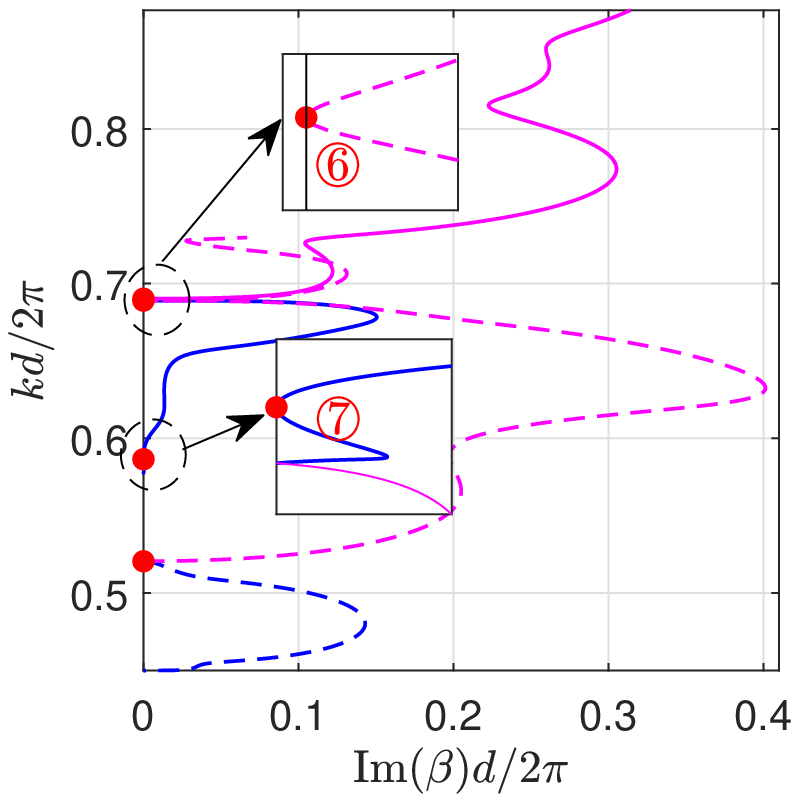}}\\
  \subfloat[]{\includegraphics[width=0.5\linewidth]{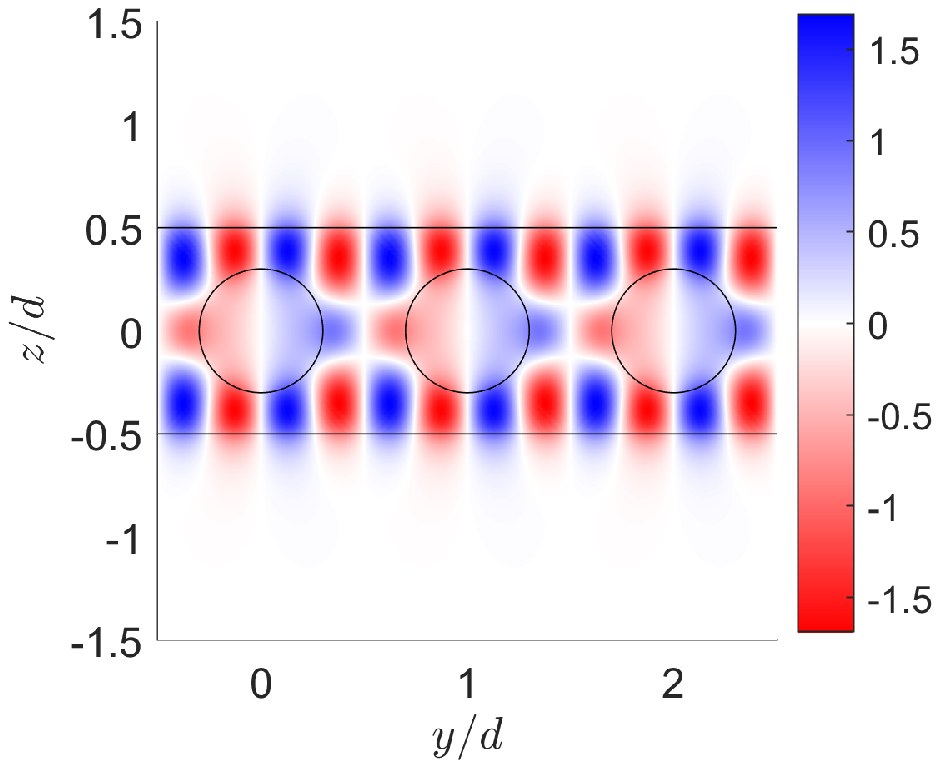}}~
  \subfloat[]{\includegraphics[width=0.5\linewidth]{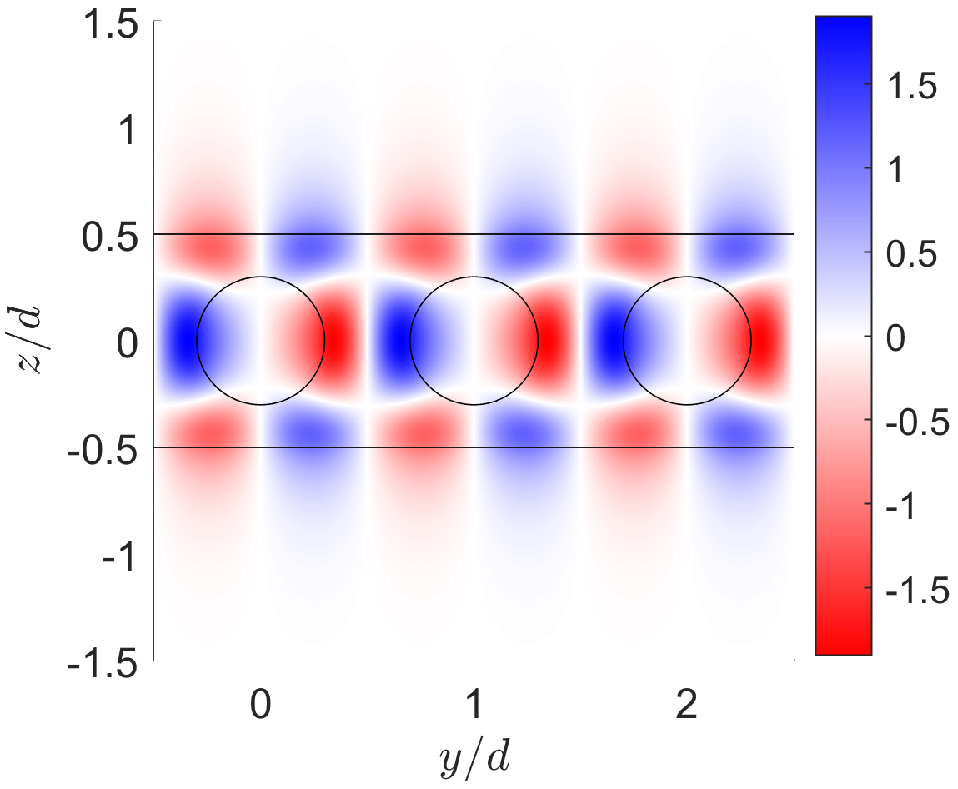}}\\
  \subfloat[]{\includegraphics[width=0.5\linewidth]{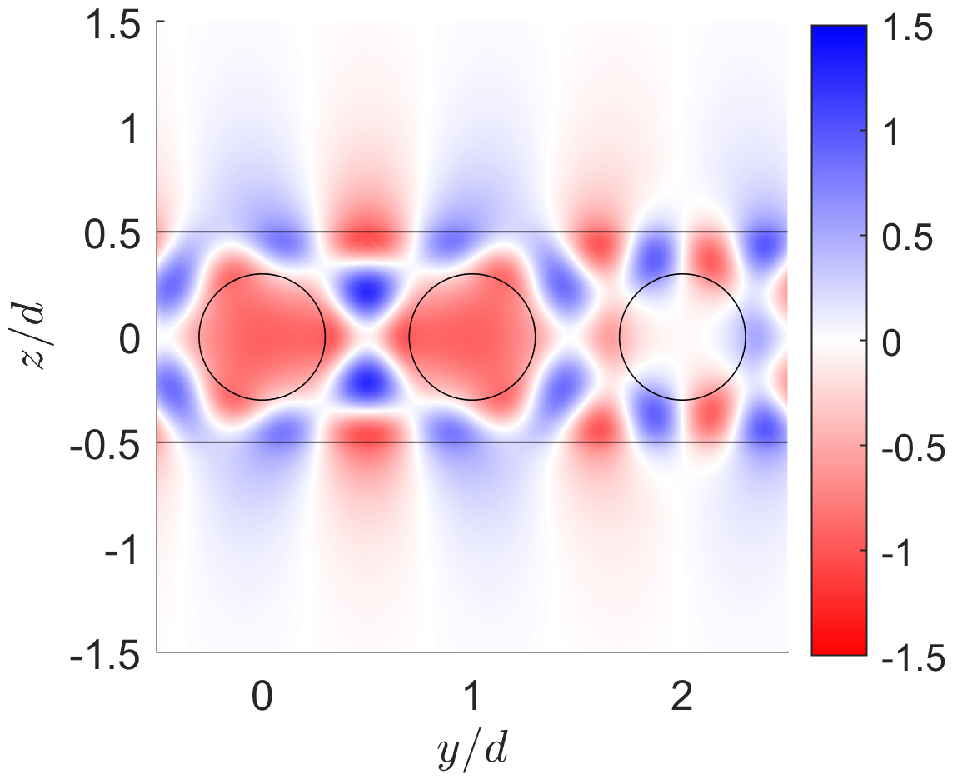}}~
  \subfloat[]{\includegraphics[width=0.5\linewidth]{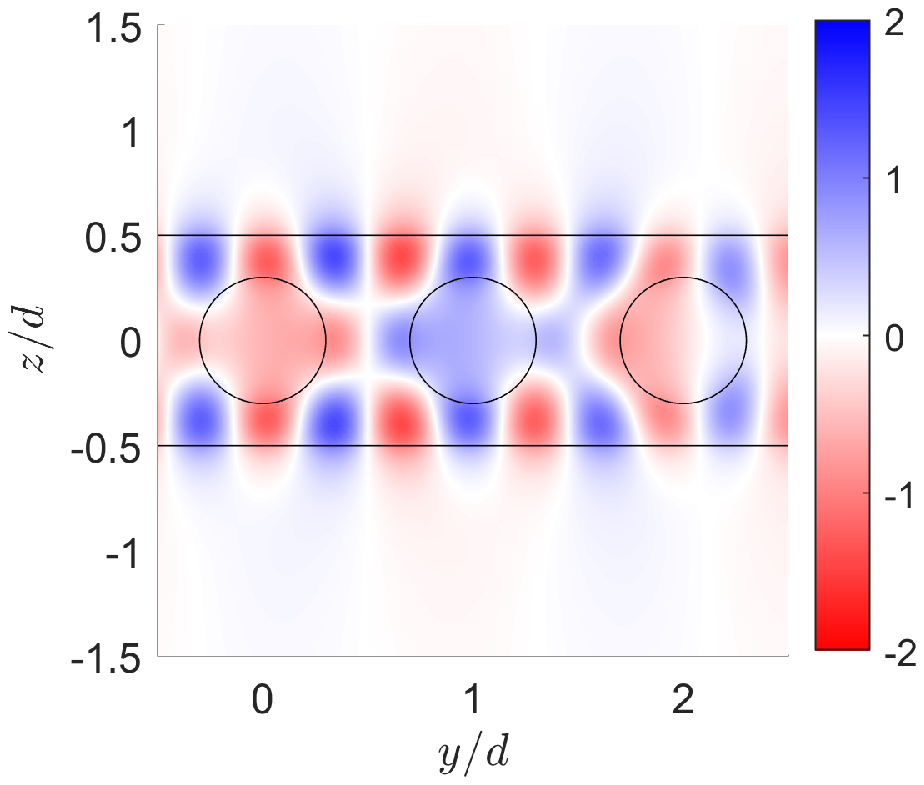}}  
  \caption{Example 2: a slab with a periodic array of air holes. (a,b)
    Dispersion curves of complex (blue) and leaky (purple) modes, with
    BICs shown as the red dots: (a)
    $k$ versus $\RE(\beta)$, (b) $k$ versus $\Im(\beta)$.
    (c-f): Field profiles of BICs: (c) BIC ${\small \textcircled{\raisebox{-.9pt}4}}$, (d)
    BIC ${\small \textcircled{\raisebox{-.9pt}5}}$, (e) 
    BIC ${\small \textcircled{\raisebox{-.9pt}6}}$, (f) 
    BIC ${\small \textcircled{\raisebox{-.9pt}7}}$.}
  \label{Fig5}
\end{figure}

The second example is a slab with a periodic array of air holes,  as shown 
Fig.~\ref{Fig1}(b). The parameters are $\vare_1 = 1$, $\vare_2 = 11.56$, $a = 0.3d$
and $h = d$. Like the first example, this periodic structure supports
a few BICs. In Fig.~\ref{Fig5}(a), 
four BICs are shown as the red dots and they are marked by
${\small \textcircled{\raisebox{-.9pt}4}}$, 
${\small \textcircled{\raisebox{-.9pt}5}}$, 
${\small \textcircled{\raisebox{-.9pt}6}}$
and 
${\small \textcircled{\raisebox{-.9pt}7}}$, respectively. 
BICs ${\small \textcircled{\raisebox{-.9pt}4}}$ and 
${\small \textcircled{\raisebox{-.9pt}5}}$ are anti-symmetric standing
waves. Their frequencies are $\omega_* = 0.6902 (2\pi c/d)$ and
$0.5204 (2\pi c/d)$, respectively. 
The other two BICs are propagating BICs with a nonzero $\beta_*$. BIC
${\small \textcircled{\raisebox{-.9pt}6}}$ has Bloch wavenumber
$\beta_* = 0.1632 (2\pi/d)$ and frequency 
$\omega_* =  0.6890 (2\pi c/d)$. For BIC
${\small \textcircled{\raisebox{-.9pt}7}}$, we have
$\beta_* = 0.3829 (2\pi/d)$ and $\omega_* = 0.5864 (2\pi c/d)$.

As predicted by the theory developed in Sec.~III(B), for each anti-symmetric
standing wave, a leaky mode and a {\it complex mode} emerge at $\beta=0$ for
$k > k_*$ and $k < k_*$, respectively. The {\it complex mode} emerged from
BIC ${\small \textcircled{\raisebox{-.9pt}5}}$ ends at the maximum
point on the dispersion curve of a regular guided mode below the light
line~\cite{cmode20}. 
The {\it complex mode} emerged from
BIC ${\small \textcircled{\raisebox{-.9pt}4}}$ turns to a leaky mode
at a transition point with a real $\beta$. This transition point
corresponds to a special diffraction solution with incident wave from
one diffraction channel and outgoing wave in a different radiation
channel~\cite{cmode20}.  For the leaky and {\it complex} modes emerged from ${\small
  \textcircled{\raisebox{-.9pt}4}}$, the real and imaginary parts of
$\beta$ have complicated dependence on $k$. The propagating BIC
${\small \textcircled{\raisebox{-.9pt}6}}$
lies on the dispersion curve of the leaky mode emerged from BIC
${\small \textcircled{\raisebox{-.9pt}5}}$. Consistent with the theory
in Sec.~III(A), this BIC has a positive power and the derivative
$d\beta/dk$ is positive at $k_*$. The propagating BIC
${\small \textcircled{\raisebox{-.9pt}7}}$
appears on the dispersion
curve of the {\it complex mode} emerged from
BIC
${\small \textcircled{\raisebox{-.9pt}4}}$.
Since $d \beta/dk$ is negative at $k_*$, BIC ${\small
  \textcircled{\raisebox{-.9pt}2}}$ has a negative power, consistent
with the theory of Sec.~III(A).


\section{Conclusion}
\label{S5}

In periodic structures, a BIC is often considered as a special
state in a band of resonant modes with a real Bloch wavevector and a
complex frequency, but  for optical waveguides, eigenmodes are often
studied for a given real frequency. In this paper, we showed that a
BIC in a periodic waveguide is a special guided mode in a band of
waveguide modes with a complex Bloch wavenumber $\beta$. While the
complex-frequency modes near a BIC are all resonant modes radiating
out power laterally, the \textcolor{black}{waveguide modes with a
  complex $\beta$} can be leaky 
modes that radiate out power laterally or {\it complex modes} that decay
exponentially in the lateral direction. These two cases are simply
determined by the sign of the power carried by the BIC. If the BIC
carries no power, as in the case of standing waves, both leaky and
{\it complex} modes appear  for frequencies near the frequency of the BIC. 

Our study provides a \textcolor{black}{useful guidance} for applications of BICs in
\textcolor{black}{periodic} optical waveguides. For simplicity, we studied only eigenmodes of $E$
polarization in 2D 
structures with a single periodic direction. Our theory can be 
extended to other wave-guiding structures with BICs, such as fibers with a
periodic Bragg grating~\cite{fiber19}, periodic arrays of spheres or
disks~\cite{bulg17prl,sadbel19}, and 
uniform optical waveguides with lateral leaky
channels~\cite{zou15,lijun21oe}. The current 
work is limited to generic cases so that $\IM(\beta)$ satisfies
Eq.~(\ref{scaling1}) or (\ref{scaling2}) for BICs 
with nonzero or zero power, respectively. It is probably useful to analyze
non-generic BICs for which $\IM(\beta)$ exhibits higher order relations
with the frequency difference. 

\section*{Acknowledgments}
The authors acknowledge support from the Research Grants Council of
Hong Kong Special Administrative Region,  China (Grant No. CityU 11307720).

\section*{Appendix}
\renewcommand{\theequation}{A\arabic{equation}}
\setcounter{equation}{0}

To find $\beta_1$ for subsection A of Sec.~III, we multiply Eq.~(\ref{Eqq2}) by
$\overline{\phi}_*$ and integrate on $\Omega$. Since $\phi_*$ satisfies $\mathcal{L} \phi_*=0$, standard integration
by parts gives 
$\int_\Omega \overline{\phi}_* \mathcal{L}\phi_1 d{\bm  r} = 0$. 
Therefore, 
$\mathcal{P} \beta_1 = \int_\Omega \vare |\phi_*|^2
d{\bm r}$, 
where
\[
  \mathcal{P} = 2 \int_\Omega  \overline{\phi}_* ( \beta_* \phi_* - i
  \partial_y \phi_*)   d{\rm r}.
\]
Since $\int_\Omega \partial_y (\overline{\phi}_* \phi_*) d{\bm r} =
0$,  $\int_\Omega \overline{\phi}_*  \partial_y \phi_*  d{\bm r}$ is
pure imaginary and thus $\mathcal{P}$ is real. 
Since $u_* = \phi_* e^{ i \beta_* y}$, $\mathcal{P}$ is also given in Eq.~(\ref{defP}). The
power in the $y$ direction carried by the BIC  is
\begin{equation}
  P_*  = \frac{1}{2 Z_0 k_*} \int_{-\infty}^\infty \IM (\overline{u}_*
  \partial_y u_*) dz,   
\end{equation}
where $Z_0$ is the free space  impedance, and it is independent of
$y$. Therefore, $\mathcal{P} = 4d Z_0 k_* P_*$.

Multiplying Eq.~(\ref{Eqq4}) by $\overline{\phi}_*$ and integrating on
$\Omega$, we get
\begin{equation}
  \label{beta2}
  \mathcal{P} \beta_2 +
  \beta_1^2 \int_\Omega |\phi_*|^2 d{\bm r} +  \int_\Omega  R({\bm r})
  d{\bm r} = 0, 
\end{equation}
where $R({\bm r}) = \overline{\phi}_* [ 2 \beta_* \beta_1 \phi_1 - 2
  i  \beta_1 \partial_y \phi_1 - \varepsilon({\bm r}) \phi_1 ]$. It is
  easy to show that
  \[
    \int_\Omega R({\bm r}) d{\bm r} = \int_\Omega  \phi_1
    \overline{G}({\bm r}) d{\bm r}
    = \int_\Omega \phi_1 \overline{ \mathcal{L} \phi_1} d{\bm r}, 
  \]
  where $G({\bm r})$ is the right  hand side of Eq.~(\ref{Eqq3}).
  Therefore,
  \begin{equation}
    \label{imbeta2}
    \mathcal{P} \,    \IM(\beta_2)  = -  \IM \int_{\Omega} \phi_1 \overline{
  \mathcal{L} \phi_1} d{\bm r}.
\end{equation}
If $\phi_1$ has the far field expression
\[
  \phi(y,z) \sim b_0^\pm e^{ i \alpha_* z}, \quad z \to \pm \infty,
\]
then we can show that 
\[
\IM \int_{\Omega} \phi_1 \overline{
  \mathcal{L} \phi_1} d{\bm r}
= -d \alpha_* ( |b_0^+|^2 + |b_0^-|^2).
\]
Therefore,
\begin{equation}
  \label{imbeta2a}
  \IM(\beta_2) = \frac{ d \alpha_* ( |b_0^+|^2 + |b_0^-|^2)}{\mathcal{P}}.
\end{equation}

The functions $\psi_1$ and $\psi_2$ are related to diffraction
solutions $w_1$ and $w_2$ by Eq.~(\ref{defpsi}), and they have the
following far field expressions
\begin{eqnarray*}
  & \psi_1({\bm r}) \sim e^{i \alpha_* z} + R_1 e^{- i \alpha_* z}, \quad
  & z                       \to -\infty, \\
&  \psi_1({\bm r}) \sim T_1 e^{i \alpha_* z}, \quad & z  \to +\infty, \\
&  \psi_2({\bm r}) \sim T_2 e^{-i \alpha_* z}, \quad &z  \to -\infty, \\
&  \psi_2({\bm r}) \sim e^{-i \alpha_* z} + R_2 e^{i \alpha_* z}, \quad &z  \to +\infty,
\end{eqnarray*}
where $R_1$, $R_2$, $T_1$ and $T_2$ are the reflection and
transmission coefficients. Using these asymptotic expressions, we can
calculate $F_1$ and  $F_2$  satisfying 
\[
  F_j = \int_\Omega \overline{\psi}_j G({\bm r}) d{\bm r} =
  \int_\Omega \overline{\psi}_j \mathcal{L} \phi_1  d{\bm r}.
\]
The result can be written as
\begin{equation}
  \label{eq:2}
  \begin{bmatrix}
    F_1 \cr F_2 
  \end{bmatrix}
  = 2i d \alpha_* \overline{S}
  \begin{bmatrix}
    b_0^- \cr b_0^+
  \end{bmatrix},
  \quad S =
  \begin{bmatrix}
    R_1 & T_2 \cr T_1 & R_2
  \end{bmatrix}.
\end{equation}
The scattering matrix $S$ is unitary. Therefore,
\[
  |F_1|^2 + |F_2|^2 = 4d^2 \alpha_*^2 ( |b_0^+|^2 + |b_0^-|^2). 
\]
Inserting the above into Eq.~(\ref{imbeta2a}), we get
Eq.~(\ref{beta2new}).

For subsection B of Sec.~III, the functions $G$ and $R$ are defined differently. 
For  $R({\bm r})$ given after Eq.~(\ref{Eqq12}), it is easy to
show that
\[
  \int_\Omega R({\bm r}) d{\bm r}
  = \int_\Omega \hat{\phi}_1 \overline{G}({\bm r})  d{\bm r}
  =  \int_\Omega \hat{\phi}_1 \overline{ \mathcal{L} \hat{\phi}_1 }
  d{\bm r}
\]
where $G$ is given in Eq.~(\ref{hatphi1}). Following the same steps
above, we get 
\begin{equation}
  \label{f1f2n}
  \IM \int_\Omega \hat{\phi}_1 \overline{ \mathcal{L} \hat{\phi}_1 }
  d{\bm r}
  = - \frac{ |F_1|^2 + |F_2|^2}{ 4 d \alpha_*},
\end{equation}
where $F_1$ and $F_2$ are given by
\[
  F_j = \int_\Omega \overline{\psi}_j G({\bm r})  d{\bm r} =
  \int_\Omega \overline{\psi}_j \mathcal{L} \hat{\phi}_1  d{\bm r}.
\]
The above leads to Eq.~(\ref{rtof}).

\end{document}